\documentclass[a4paper,11pt]{article}
\usepackage{pos}

\title{Four-quark states from functional methods}

\author*[a]{Joshua Hoffer}
\author[a,b]{Christian S. Fischer}

\affiliation[a]{Institut für Theoretische Physik, Justus-Liebig-Universität Gießen\\
  35392 Gießen, Germany}

\affiliation[b]{Helmholtz Forschungsakademie Hessen für FAIR (HFHF), GSI Helmholtzzentrum für Schwerionenforschung, Campus Gießen\\
35392 Gießen, Germany}

\emailAdd{joshua.hoffer@theo.physik.uni-giessen.de}
\emailAdd{christian.fischer@theo.physik.uni-giessen.de}

\abstract{The discovery of four-quark states attracted a lot of attention from the theoretical as well as the 
	experimental side. To study their properties from QCD we use a functional framework which combines (truncated) 
	Dyson-Schwinger and Bethe-Salpeter equations in Landau gauge. This approach allows us to extract qualitative 
	results for mass spectra, decay widths and wavefunctions of candidates for bound as well as resonant 
	four-quark states. Furthermore, we can investigate the possible internal structure of such states. 
	We report on recent developments and results using this functional framework and give an overview 
	about the current status as well as future developments.}

\FullConference{%
  FAIR next generation scientists - 7th Edition Workshop (FAIRness2022)\\
  23-27 May 2022\\
  Paralia (Pieria, Greece)
}

\begin{document}
\maketitle

\section{Functional Framework}
To study the properties of hadrons we employ a non-perturbative functional framework in which we combine Dyson-Schwinger equations (DSEs), i.e., the quantum equations of motion, with hadronic bound state equations, e.g., Bethe-Salpeter equations (BSEs) (see \citep{Eichmann2016} for a detailed review and references therein). The functional approach has been successfully applied to the meson and baryon spectrum (see, e.g., \citep{Eichmann2016}), glueballs \citep{SanchisAlepuz2015,Huber2020} and also the spectrum of light and heavy-light four-quark states \citep{Eichmann2016a,Wallbott2019,Wallbott2020,Eichmann2020,Santowsky2022,Santowsky2022a}.

A BSE can be thought of as an eigenvalue equation
\begin{equation}\label{eq: homog.BSE}
\lambda(P^2)\, \Gamma = K G\, \Gamma\, ,
\end{equation}
where $\Gamma$ is the Bethe-Salpeter amplitude (BSA), $K$ denotes the interaction kernel and $G$ are the fully dressed quark propagators. We have also introduced an eigenvalue $\lambda(P^2)$ which depends on the hadron momentum $P$ squared. Eq.~(\ref{eq: homog.BSE}) is solved if $\lambda(P^2 = -M^2) = 1$, i.e., the hadron goes on-shell.

\begin{figure}
\centering
\includegraphics[scale=0.5]{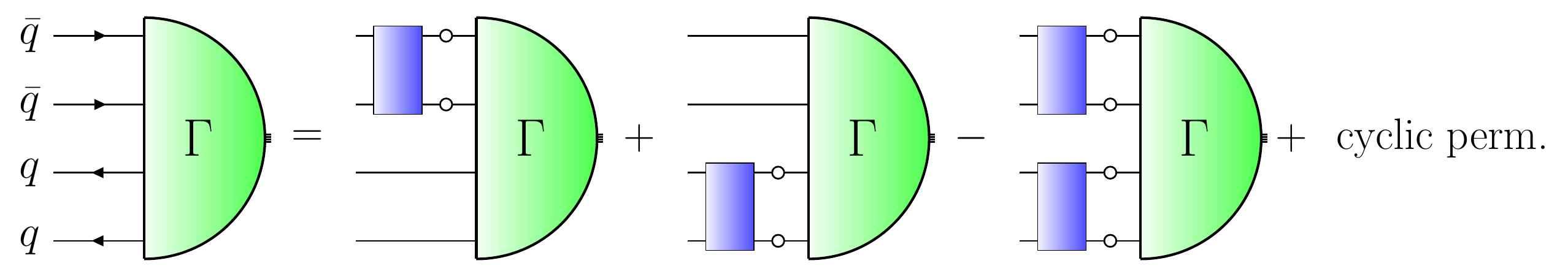}
\caption{Four-quark BSE with two-body interactions. The green half-circles denote the BSA and the blue boxes indicate interactions between two quarks. There are three interaction topologies: two meson-meson ($q\bar{q}$)($q\bar{q}$) topologies and one a diquark-antidiquark ($qq$)($\bar{q}\bar{q}$) topology (shown).}
\label{fig:4quarkBSE}
\end{figure}

In the following we will focus on the properties of four-quark states. Four-quark states are systems of two quarks and two anti-quarks, i.e., $qq\bar{q}\bar{q}$. 
Their exact four-body Bethe-Salpeter equation contains irreducible two-, three- and 
four-body interactions. For reasons of complexity, the latter two have been neglected 
so far. While on the surface this may be considered an uncontrolled and potentially 
severe oversimplification, arguments in favour of the dominance of two-body interactions 
have been discussed in \citep{Eichmann2016a}. We are then left with the system shown 
in Fig.~\ref{fig:4quarkBSE}, where we have three different two-body interaction topologies: two meson-meson and one diquark-antidiquark. For the two-body interaction kernels, we employ the \textit{rainbow-ladder truncation}, i.e., the interaction reduces to exchanges of effective gluons, for details see, e.g., \citep{Maris1997,Maris1999}. In recent years there were some developments to systematically improve the Ansatz for the interaction in the light meson sector \citep{Heupel2014,Williams2016}.\\
The four-body BSA for the scalar four-quark state has the following form
\begin{equation}
\Gamma(k,q,p,P) = \sum_{i=1}^N\, f_i(\Omega)\, \tau_i(k,q,p,P) \otimes \Gamma_C\otimes \Gamma_F\, ,
\end{equation}
where $k,q$ and $p$ are relative momenta between (anti-)quark pairs, each associated with one specific interaction topology, and $P$ is the total momentum of the four-quark state. The $f_i$ are the dressing functions of the four-quark state, which depend on a set of nine Lorentz-invariants $\Omega$ and $\tau_i$ are the respective basis elements in Dirac space. For the scalar four-quark state we have $N=256$ basis elements. Furthermore, there is also a colour and a flavour part, $\Gamma_C$ and $\Gamma_F$ respectively.
We apply two different strategies to make this equation more manageable to solve. 
First, we consider the 16 $s$-wave tensors, which depend only on $P$ but not on the relative momenta and form a Fierz-complete basis. This approximation has been done in the three-body equation for baryons and it was found to be reliable to $\sim 10\%$, which is sufficient for the current purposes. The second strategy is to recast the momenta $k,q,p,P$ into multiplets of the permutation group $S_4$ \citep{Eichmann2015}. One can then identify a singlet $S_0$, a doublet $D$ and two triplets $T_0,$ $T_1$. 
It has been shown in \citep{Eichmann2016a}, that the dependence of the dressing functions $f_i$ on the triplet variables 
is weak and can be neglected. The other two, however, are important: the $S_0$-variable carries the scale, whereas $D$ 
restricts the phase space. It is also in the doublet-variables $D$ that intermediate 2-body pole structures arise dynamically.
As a further approximation we use a physically motivated basis, i.e., we assume that the amplitude is dominated either by 
tensor structures corresponding to two physical channels identified by the decay products with lowest mass or by the diquark 
channels with lowest (unphysical) mass.
Thus, we have three dressing functions $f^{\mathcal{M}_1},\,f^{\mathcal{M}_2},\, f^{\mathcal{D}}$ for the two meson and the diquark topology and can now put in the pole structures by hand, i.e., we make the replacement $f_i(S_0,D) \to f_i(S_0)\cdot P_{ab} \cdot P_{cd}$,
where $P_{ab/cd}$ denote 2-body poles in certain topologies and $a,b,c,d$ is the quark index. In this physical basis, the dressing functions $f_i$ only depend on the singlet $S_0$. A more detailed description can be found in \citep{Eichmann2020}.
Note, there is also a related BSE approach in which one assumes dominant two-body forces and can then further simplify the four-quark BSE. This approach will be termed \textit{two-body approach} and deals with effective meson-meson and diquark-antidiquark degrees of freedom, which interact via quark exchange, see \citep{Eichmann2020,Santowsky2022a} for details. 

\section{Results}
In previous works, the masses $M^2 = -P^2$ of the light and heavy-light four-quark states were extracted in the 
four-body and the two-body approaches, see \citep{Eichmann2016a,Wallbott2019,Wallbott2020,Santowsky2022a}.
As most hadrons are resonances one strives to extract the real part of the mass plus the decay width of 
the hadrons from the theory framework to compare with the experimental values. In principle, it is straight 
forward (but numerically tedious) to solve the BSE, Eq.~(\ref{eq: homog.BSE}), for such complex $P^2$. 
In the rest frame of the hadron this corresponds to $P^\mu = (i M + \Gamma/2)\, \hat{e}_4^\mu$ and
the complex eigenvalue has to fulfil two conditions: $\text{Re}(\lambda(P^2)) = 1$ and $\text{Im}(\lambda(P^2)) = 0$. 

\begin{figure}
\includegraphics[scale=0.5]{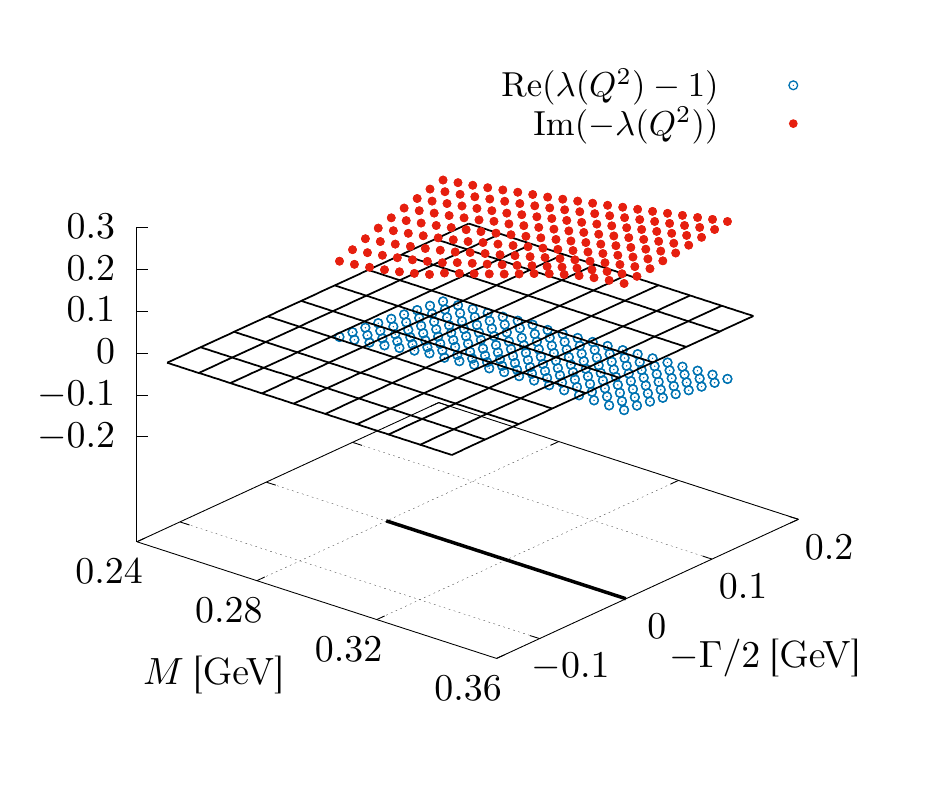}\includegraphics[scale=0.5]{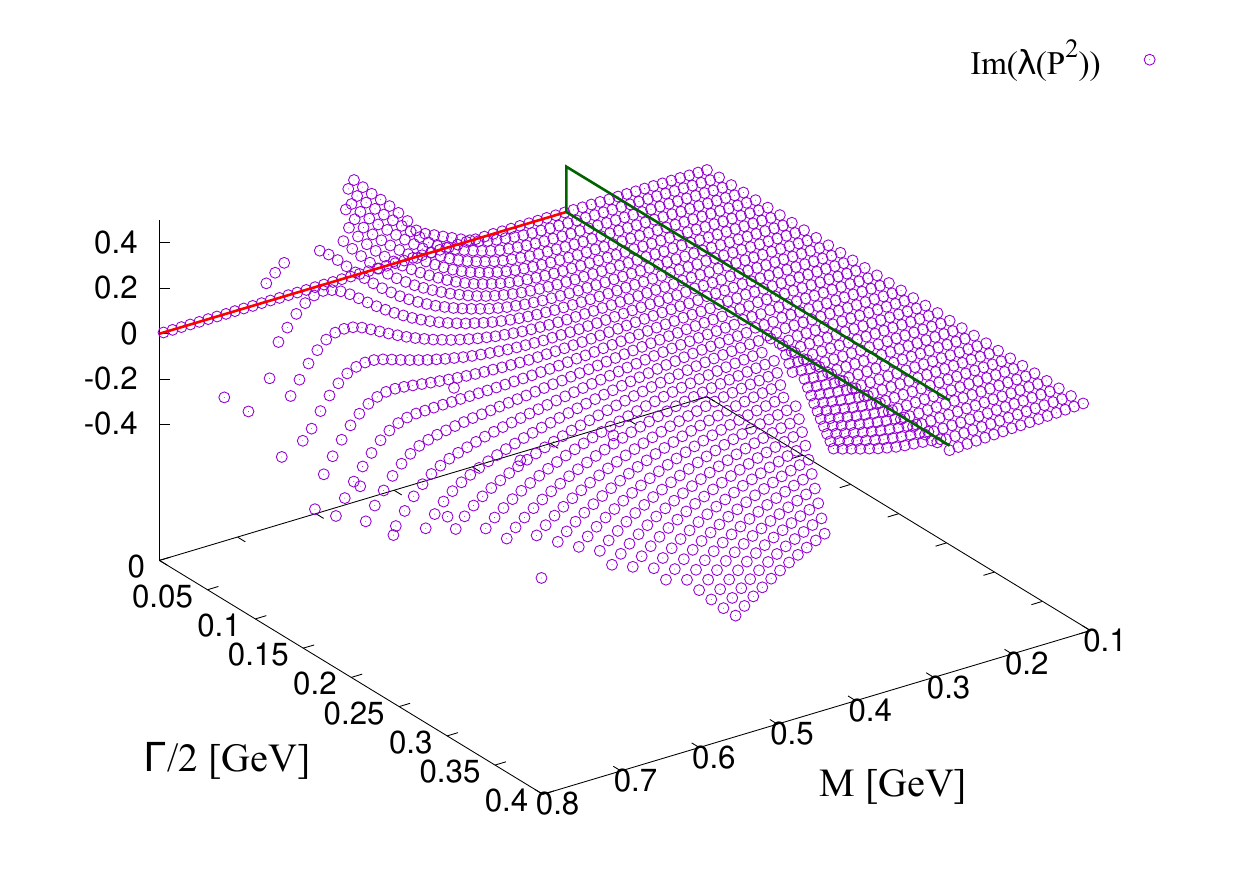}\includegraphics[scale=0.28]{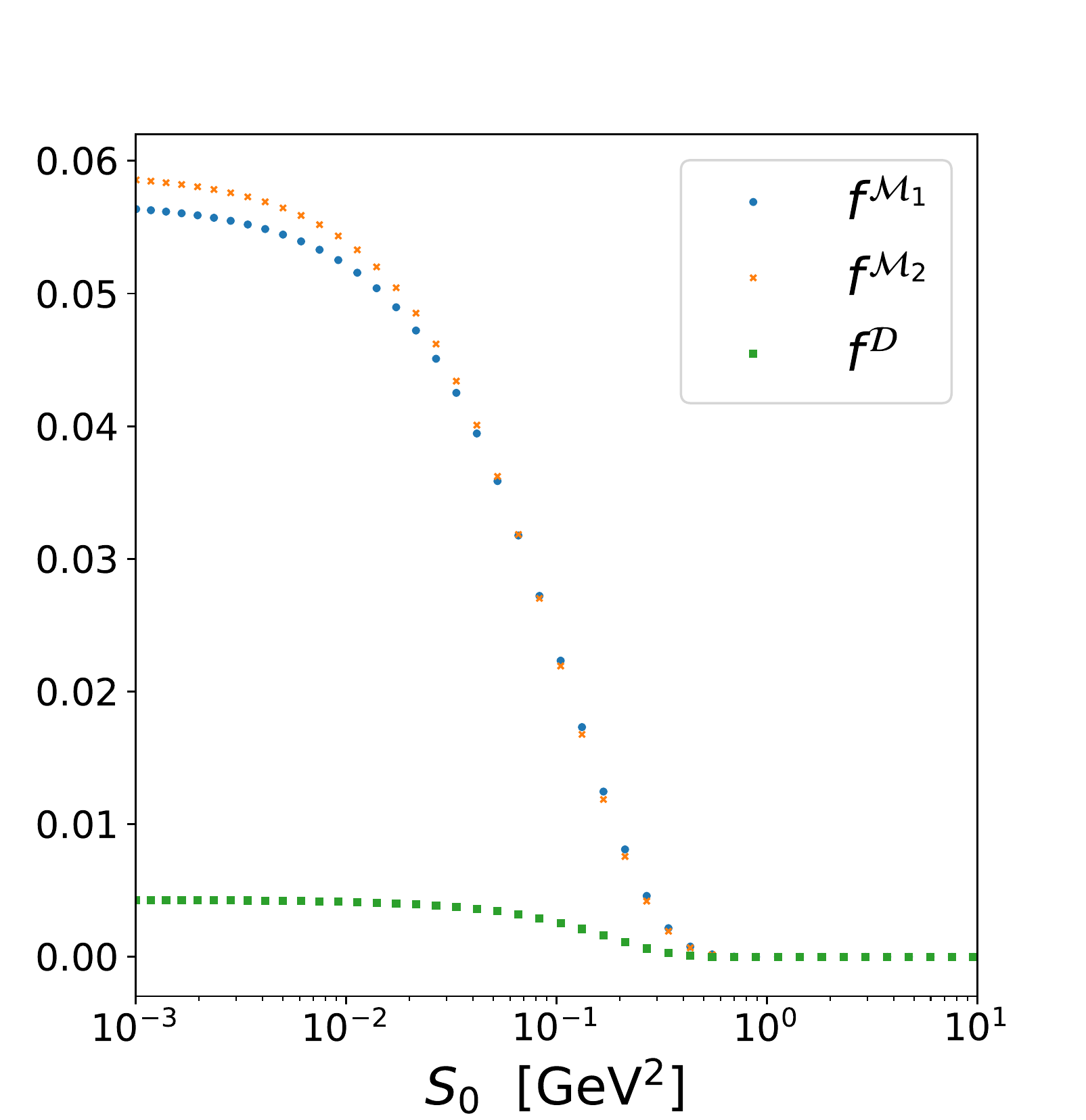}
\caption{\textit{Left:} Real and Imaginary part of the eigenvalue $\lambda(Q^2)$ for the $\sigma$ in the two-body formalism. Here the total momentum is denoted by $Q^2$. The black line on the bottom indicates the branch cut along the real axis. Figure taken from \citep{Santowsky2022}. \textit{Middle:} Preliminary result for the imaginary part of $\lambda$ for the $\sigma$ in the four-body formalism. The red line is the branch cut along the real axis and the green box is the particle threshold. \textit{Right:} Preliminary result showing the dressing functions for the $a_0$ as a function of the singlet variable $S_0$. From the magnitude, we find that the meson contributions are dominant and the diquark contribution is subleading.}
\label{Fig:Eigenvalue}
\end{figure}

There are however two caveats. With the techniques available so far, the computation is only possible in the 
first (unphysical) Riemann sheet. As we are interested in resonances, which are identified by poles lying in 
the (physical) second Riemann sheet, we need to perform an analytic continuation to extract the physical results. 
The second caveat is, that it is sometimes not straight forward to reach the $P^2$ such that $\lambda(P^2=-M^2) = 1$ 
is fulfilled. This is because of the intermediate particle poles, which introduce branch cuts into the equation. 
Fortunately, in many cases path deformation techniques are available to circumvent the problem. These techniques 
have been explored for conventional mesons, see, e.g., \citep{Eichmann2020} and references therein, and also in 
the context of four-quark states in the two-body approach \citep{Santowsky2022}.
In this paper, the authors used a combination of path deformation and analytic continuation in the two-body approach to 
compute the BSE in the complex plane above the threshold, see left plot in Fig.~\ref{Fig:Eigenvalue}. It shows the 
real and imaginary part of the eigenvalue $\lambda(P^2)$ for the $\sigma$ meson and one can clearly see a branch 
cut opening in the imaginary part of $\lambda(Q^2)$ above a certain decay threshold, i.e., $\pi\pi$ for the $\sigma$. 
Extrapolating beyond the cut, they extracted the masses plus the decay widths of the light scalar nonet particles 
$\sigma,\, a_0,\, f_0$. 
The results match reasonably well with experimental values (see \citep{Santowsky2022} for a full discussion).

The path deformation technique, however, has not yet been exploited in the four-body approach. This is work in progress. 
On the right in Fig.~\ref{Fig:Eigenvalue}, we show a preliminary result for the imaginary part of the eigenvalue of 
the $\sigma$ obtained in the four-body formalism. We have identified the branch cut opening above the two-pion 
threshold, similar to the two-body approach, and are now in the process of adapting the path deformation techniques to 
the problem at hand. 

In the regions of $P^2$ which are directly assessable, we are now also in a position to address the internal structure 
of the four-quark states directly via a comparison of the size of different contributions to the Bethe-Salpeter amplitude, 
i.e. the shape and magnitude of the dressing functions $f^{\mathcal{M}_1},\,f^{\mathcal{M}_2},\, f^{\mathcal{D}}$. 
On the right in Fig.~\ref{Fig:Eigenvalue} we plotted the dressing functions for the example case of the $a_0$ as functions 
of $S_0$, i.e., the overall momentum scale. We find a clear dominance of the meson dressing functions over the diquark 
dressing function. This is completely in line with the findings from the two-body approach, \citep{Santowsky2022}, where 
this information was only indirectly available.

\paragraph{Closing remarks} Since the DSE/BSE framework in principle makes no assumptions on the internal structure of 
four-quark states (i.e., e.g., diquark clustering vs. meson clustering), it is a very interesting tool to study the dynamical 
effects which generate these structures. For all states studied so far, the (heavy-light or light-light) meson-meson 
components dominate and diquark components are subleading. In recent time, the framework evolved to a level, which makes 
the extraction of decay widths possible \citep{Santowsky2022}, at least in the two-body approach. The generalisation 
of the corresponding techniques to the four-body approach is work in progress 
and may lead to more refined statements about the internal structure of four-quark states.
\section{Acknowledgments}
This work was supported by HGS-HIRe for FAIR, the GSI Helmholtzzentrum für Schwerionen-
forschung, and the BMBF under project number 62001011. We acknowledge computational resources
provided by the HPC Core Facility and the HRZ of the Justus-Liebig-Universität Gießen.

\bibliographystyle{JHEP}
\bibliography{fairnessbib}

\end{document}